\documentclass[12pt]{article}
\pdfoutput=1
\usepackage{putex}
\usepackage{comment}
\usepackage{graphicx}
\usepackage{caption}
\usepackage{amsmath}
\usepackage[usenames,dvipsnames,table]{xcolor}
\usepackage{array}
\usepackage{subcaption}
\usepackage{epstopdf}
\usepackage{enumerate}
\usepackage{cite}
\usepackage{tensor}
\usepackage{slashed}
\usepackage[export]{adjustbox}
\usepackage[aligntableaux=center]{ytableau}
\usepackage{dsfont}
\usepackage[utf8]{inputenc}
\usepackage[
      colorlinks=true,
      linkcolor=blue,
      urlcolor=blue,
      filecolor=black,
      citecolor=red,
      ]{hyperref}
\newcommand{\RNum}[1]{\uppercase\expandafter{\romannumeral #1\relax}}

\newcommand {\be} {\begin {equation}}
\newcommand {\ee} {\end {equation}}

\newcommand {\bes} {\begin {equation*}}
\newcommand {\ees} {\end {equation*}}

\newcommand{\beq}{\begin{equation}}
\newcommand{\eeq}{\end{equation}}

\def\ie{\begin{equation}\begin{aligned}}
\def\fe{\end{aligned}\end{equation}}

\numberwithin{equation}{section}

\def\<{\langle}
\def\>{\rangle}

\begin{document}

\preprint{PUPT-2644}


\institution{PU}{ $^{a}$ Department of Physics, Princeton University, Princeton, NJ 08544, USA }
\institution{}{ $^b$ Center for Cosmology and Particle Physics, New York University, New York, NY 10003, USA}

\title{Line Defects in Fermionic CFTs } 

\authors{Simone Giombi$^a$, Elizabeth Helfenberger$^a$, and Himanshu Khanchandani$^b$}

\abstract{We study line defects in the fermionic CFTs in the Gross-Neveu-Yukawa universality class in dimensions $2<d<4$. These CFTs may be described as the IR fixed points of the Gross-Neveu-Yukawa (GNY) model in $d=4-\epsilon$, or as the UV fixed points of the Gross-Neveu (GN) model, which can be studied using the large $N$ expansion in $2<d<4$. These models admit natural line defects obtained by integrating over a line either the scalar field in the GNY description, or the fermion bilinear operator in the GN description. We compute the beta function for the defect RG flow using both the epsilon expansion and the large $N$ approach, and find IR stable fixed points for the defect coupling, thus providing evidence for a non-trivial IR DCFT. We also compute some of the DCFT observables at the fixed point, and check that the $g$-function associated with the circular defect is consistent with the $g$-theorem for the defect RG flow.}

\date{}
\maketitle

\tableofcontents

\section{Introduction and summary}

The Gross-Neveu (GN) model is a theory of interacting fermions described by the following action \cite{PhysRevD.10.3235}
\begin{equation} \label{ActionGN}
S = - \int d^d x \left(  \bar{\Psi}_i \gamma \cdot \partial \Psi^i + \frac{g}{2} \left( \bar{\Psi}_i \Psi^i \right)^2  \right)
\end{equation}
where the index $i$ runs from one to $N_{f}$, the number of fermion flavors, so we have $N_f$ Dirac fermions. It was originally introduced as a toy model for asymptotic freedom. In $d = 2 + \epsilon$, the beta function for the coupling $g$ in (\ref{ActionGN}) has a UV fixed point where $g$ is of order $\epsilon$. As we shall review below, the Gross-Neveu model admits a ``UV completion" in terms of a Gross-Neveu-Yukawa theory \cite{Hasenfratz:1991it, Zinn-Justin:1991ksq}, which includes $N_f$ fermions and one real scalar field. Its upper critical dimension is $d=4$, and the model has IR stable fixed points in $d=4-\epsilon$. Thus, one expects a family of fermionic CFTs in the range $2<d<4$, which we may call Gross-Neveu-Yukawa or Gross-Neveu CFTs. In particular, one expects interacting CFTs in this universality class in the physical dimension $d=3$.   This model has been extensively studied using traditional methods (see \cite{Moshe:2003xn, Gracey:2008mf,  Fei:2016sgs, Goykhman:2020tsk} and references therein), and also modern conformal bootstrap methods \cite{Iliesiu:2015qra, Iliesiu:2017nrv, Erramilli:2022kgp}. A version of the Gross-Neveu-Yukawa model has also been proposed to describe the semi-metal to insulator transition in the Hubbard model \cite{Herbut:2006cs, Assaad:2013xua}. More recently, there have been studies of the Gross-Neveu/Gross-Neveu-Yukawa CFTs in the presence of a boundary \cite{Carmi:2018qzm, Giombi:2021cnr, Herzog:2022jlx}. In this paper, we introduce and study line defects in these fermionic CFTs and provide evidence for the existence of a conformal line defect.

An equivalent description of the GN model (\ref{ActionGN}) may be written in terms of a Hubbard-Stratonovich field $\sigma$ 
\begin{equation}
S = -\int d^d x \left( \bar{\Psi}_i \gamma \cdot \partial \Psi^i + \frac{1}{\sqrt{N}} \sigma \bar{\Psi}_i \Psi^i - \frac{\sigma^2}{2 g N} \right)
\label{GN-HS}
\end{equation}
where we defined $N = N_f c_d$ with $c_d$ being the number of components of a Dirac fermion in $d$ dimensions. Integrating out the auxiliary field $\sigma$, one obtains the original Lagrangian \eqref{ActionGN}. However, the Hubbard-Stratonovich description is particularly useful to develop the $1/N$ perturbation theory of the critical theory in any dimension (see \cite{Moshe:2003xn, Giombi:2016ejx} for a review). One may obtain an induced $\sigma$ two-point function by integrating out the fermions. At the critical point (i.e., in the UV limit if we use the GN description), one may drop the $\sigma^2/ 2 g N $ term, and one then finds that the $\sigma$ field has a conformal two-point function, with scaling dimension (see e.g. \cite{Fei:2016sgs})
\begin{equation}
\label{delsig-largeN}
\begin{aligned}
&\Delta_{\sigma}=1+\frac{f_1(d)}{N}+O(1/N^2)\,,\\
&f_1(d)=\frac{2^{d + 1} (d-1) \sin \left( \frac{\pi d}{2} \right) \Gamma \left(\frac{d - 1}{2} \right)}{d (d - 2)\pi^{\frac{3}{2}} \Gamma \left(\frac{d}{2} - 1 \right)}\,.
\end{aligned}
\end{equation} 
One can see that $f_1(d)$ is negative in the range $2<d<4$ (it goes to zero at the upper and lower critical dimensions $d=4$ and $d=2$ respectively). For instance, in $d=3$ one has $f_1(d=3)=-32/(3\pi^2)$.  

Therefore, it is natural to introduce a line defect by integrating $\sigma$ along a line as (we work in flat Euclidean spacetime and the coordinates are parametrized as $x = (\tau, \mathbf{x})$): 
\begin{equation} \label{ActionLargeN}
S = -\int d^d x \left( \bar{\Psi}_i \gamma \cdot \partial \Psi^i + \frac{1}{\sqrt{N}} \sigma \bar{\Psi}_i \Psi^i   \right) + h \int d \tau \sigma(\tau, \mathbf{0}).
\end{equation}
Since at large $N$ the (bulk) dimension of $\sigma$ is slightly below 1, the defect coupling constant $h$ is weakly relevant, and there is a possibility for a non-trivial IR fixed point of the defect RG flow.  We will compute the beta function of $h$ in the $1/N$ expansion, and show that indeed there is an IR fixed point where $h$ is of order one at large $N$. This provides evidence that the system (\ref{ActionLargeN}) flows to a non-trivial IR DCFT. 

  Note that the defect breaks the parity symmetry of the original theory, which acts by sending $\bar{\Psi}_i \Psi^i  \rightarrow - \bar{\Psi}_i \Psi^i$ and $\sigma \rightarrow - \sigma$. Also note that, since $\sigma$ plays the role of the fermion bilinear via the Hubbard-Stratonovich transformation (\ref{GN-HS}), in the original Gross-Neveu description (\ref{ActionGN}) the defect (\ref{ActionLargeN}) corresponds to integrating $\bar{\Psi}_i\Psi^i$ over the line. 

As mentioned above, a UV completion of the Gross-Neveu model is provided by the Gross-Neveu-Yukawa (GNY) model \cite{Hasenfratz:1991it, Zinn-Justin:1991ksq} 
\begin{equation} \label{ActionGNY}
S = \int d^d x \left( \frac{(\partial_{\mu} s)^2}{2} - \left( \bar{\Psi}_i \gamma\cdot \partial \Psi^i + g_{1} s \bar{\Psi}_i \Psi^i\right) + \frac{g_{2}}{24} s^4   \right).
\end{equation}
The beta functions of $g_1$ and $g_2$ may be computed using standard epsilon expansion methods, and one finds that the model has a perturbative IR fixed point in $d=4-\epsilon$, for general $N_f$. This fixed point is in the same universality class as the one obtained from the Gross-Neveu model, and one may check that the CFT data computed using the epsilon expansion in (\ref{ActionGNY}) and large $N$ expansion in (\ref{GN-HS}) agree in the overlapping regime of validity (see e.g. \cite{Fei:2016sgs} for a review and collection of known results). The local field $s$ in the GNY description plays the same role as $\sigma$ in the large $N$ description, however we use a different symbol because their normalizations are a priori different. At the Wilson-Fisher fixed point of (\ref{ActionGNY}), one finds in particular the scaling dimension
\begin{equation}
\Delta_s = 1-\frac{3\epsilon}{N+6}+O(\epsilon^2).
\end{equation}
Note that at large $N$ this agrees with (\ref{delsig-largeN}) expanded near $d=4$. Analogously to (\ref{ActionLargeN}), the defect in this GNY setup is introduced by integrating the field $s$ along a line. Since the (bulk) dimension of the $s$ field is slightly below 1, as in the large $N$ description one obtains a weakly relevant flow, now for small $\epsilon$ but general $N$.  The defect in the GNY description is similar to the type of line defect that occurs in the presence of a localized magnetic field in  scalar $O(N)$ model, studied in \cite{Allais:2014fqa, Cuomo:2021kfm} (see also \cite{Rodriguez-Gomez:2022gbz}) \footnote{A similar line defect in tensor models was studied recently in \cite{Popov:2022nfq}.}. Below we will compute the beta function of the defect coupling using the $\epsilon$ expansion setup, again finding evidence for a non-trivial IR DCFT. We also show that various DCFT observables computed in the $\epsilon$ expansion in $d=4-\epsilon$ and large $N$ expansions in $2<d<4$ agree with each other in the overlapping regime of validity.  To summarize our results, we present the operator dimensions of the defect operators we studied in Table \ref{TableDefectDimensions}. In addition to the scaling dimensions of the defect local operators, an interesting observable for conformal line defects is the $g$-function, defined as the normalized expectation value of the circular defect \cite{Cuomo:2021kfm, Cuomo:2021rkm} (see also \cite{Beccaria:2017rbe, Kobayashi:2018lil} for earlier work). As was proved in \cite{Cuomo:2021rkm}, RG flows localized on a line defect in a CFT$_d$ obey a $g$-theorem:\footnote{This generalizes the previously known $g$-theorem for line defects in 2d CFT \cite{Affleck:1991tk, Friedan:2003yc, Casini:2016fgb}.} the value of $\log(g)$ in the UV is larger than the one at the IR fixed point. Moreover, \cite{Cuomo:2021rkm} also defined a defect entropy function (which reduces to $\log(g)$ at fixed points) which decreases monotonically under the defect RG flow. We calculate the value of the $g$-function in our IR DCFT in the $\epsilon$-expansion and find that it is consistent with the $g$-theorem.

\begin{table}[!ht]
\centering
{\renewcommand{\arraystretch}{1}%
\begin{tabular}{|c|c|c|}
\hline
 & GNY $d = 4 - \epsilon$ & Large $N$   \\ 
\hline
 & &  \\
Leading defect scalar  & $ 1 + \frac{6}{(N + 6)} \epsilon$ & \quad $ 1 - \frac{2^{d + 2} (d-1) \sin \left( \frac{\pi d}{2} \right) \Gamma \left(\frac{d - 1}{2} \right)}{N d (d - 2)\pi^{3/2} \Gamma \left(\frac{d}{2} - 1 \right)}$ \quad \\ [10pt]
\hline
 & &  \\
Transverse spin $l$ defect scalars & \quad$ 1 + l + \frac{6 (1 - l)}{(N + 6) (1 + 2 l)} \epsilon $ \quad& $1 + l + O(\tfrac{1}{N}) $  \\ [5pt]
\hline 
 & &  \\
 $U(N)$ fundamental defect fermions & $ \frac{3}{2} + l + O(\epsilon)$ &  $ \frac{d-1}{2} + l + O(\tfrac{1}{N})$   \\ [5pt]
\hline
\end{tabular}}
\caption{The dimensions of the defect operators induced by the bulk fundamental fields in GNY and large $N$ models. $l$ represents the quantum number under $SO(d-1)$ which is the group of rotations around the defect. The scalars are in symmetric traceless representations of $SO(d-1)$ while fermions are in spin $l + 1/2$ spinor representation of $Spin(d-1)$. We are only considering here lowest twist operators for each spin. All these receive corrections from higher orders in perturbation theory in $\epsilon$ or $1/N$.}   
\label{TableDefectDimensions}
\end{table}

The rest of this paper is organized as follows: In section \ref{Sec:GNY}, we define and study the line defect in the GNY model in $d = 4 - \epsilon$, and compute the defect beta function using perturbation theory in $\epsilon$. We also compute several pieces of DCFT data and in particular use bulk equations of motions to calculate the anomalous dimensions of a tower of defect operators. In section \ref{Sec:LargeN}, we study the same defect in the large $N$ expansion in $2<d<4$. 
We conclude with some future directions in section \ref{Sec:Conclusions}. The appendices contain some technical details and comments on the description of the line defect in $d = 2 + \epsilon$ dimensions using the action (\ref{ActionGN}).

\section{Line defect in the Gross-Neveu-Yukawa description} \label{Sec:GNY}
Let us start by discussing the line defect in the Gross-Neveu-Yukawa model 
\begin{equation}
S = \int d^d x \left( \frac{(\partial_{\mu} s)^2}{2} - \left( \bar{\Psi}_i \gamma\cdot \partial \Psi^i + g_{1,0} s \bar{\Psi}_i \Psi^i\right) + \frac{g_{2,0}}{24} s^4   \right) + h_0 \int d \tau s (\tau, \mathbf{0})  .
\end{equation}
The bulk bare couplings may be expressed in terms of a renormalized coupling \cite{Karkkainen:1993ef, Fei:2016sgs}
\begin{equation}
g_{1,0} = M^{\frac{\epsilon}{2}} \left( g_1 +  \frac{(N + 6) g_1^3 }{32 \pi^2 \epsilon} + \dots \right), \hspace{0.5cm} g_{2,0} = M^{\epsilon} \left( g_2 + \dots \right).
\end{equation}
In $d = 4 - \epsilon$ there is a fixed point in the bulk at the following values of the couplings 
\begin{equation}
\begin{split}
(g_1^*)^2 &= \frac{(4\pi)^2}{ N + 6 } \epsilon \xrightarrow{\textrm{Large N}} 
\frac{(4\pi)^2}{N} \epsilon \\
g_2^* &= \frac{(4\pi)^2 \left( -N + 6  + \sqrt{N ^2 + 132 N + 36} \right)}{6(N + 6)} \epsilon \xrightarrow{\textrm{Large N}}
 \frac{12 (4\pi)^2}{N} \epsilon.
\end{split}
\end{equation}
Similarly for the defect coupling, we may express the bare coupling in terms of the renormalized coupling as 
\begin{equation}
h_0 = M^{\frac{\epsilon}{2}} \left( h + \frac{\delta h}{\epsilon} + \dots \right). 
\end{equation}

Following \cite{Allais:2014fqa, Cuomo:2021kfm} we will fix $\delta h$ by requiring that the counterterms cancel the divergences appearing in the one-point function $\langle s(x) \rangle$. At leading order in the bulk coupling \footnote{Note that the defect coupling is not taken to be small, so we need to include diagrams to all orders in the defect coupling}, we have the following diagrams 
\begin{equation}
\langle s_0(x) \rangle = \underset{\hbox{$(a)$}}{\vcenter{\hbox{\includegraphics[scale=1]{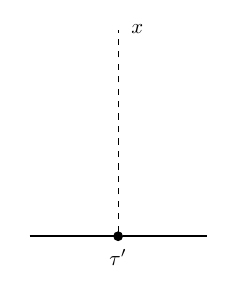}}} }\ \   + \ \  \underset{\hbox{$(b)$}}{\vcenter{\hbox{\includegraphics[scale=1]{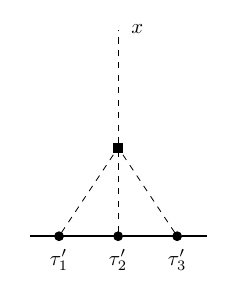}}}} \ \   + \ \  \underset{\hbox{$(c)$}}{\vcenter{\hbox{\includegraphics[scale=1]{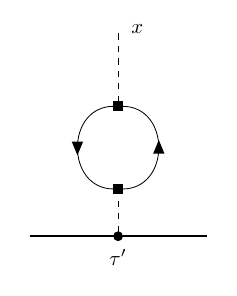}}}}
\end{equation} 
where 
\begin{equation}
\begin{split}
(a) &= - h_0 \int d \tau' \langle s(x) s (\tau', \mathbf{0}) \rangle = - \frac{h_0 \Gamma \left(\frac{d}{2} - 1 \right)}{4 \pi^{d/2}}\int d \tau' \frac{1}{\left(\mathbf{x}^2 + (\tau' - \tau)^2 \right)^{\frac{d}{2} - 1}} \\
&= - \frac{h_0  \Gamma \left(\frac{d-3}{2}\right)}{4 \pi^{(d-1)/2}  \left(\mathbf{x}^2\right)^{(d-3)/2}} =  - \frac{h_0}{4 \pi |\mathbf{x}|}
\end{split}
\end{equation}
\begin{equation}
\begin{split}
(b) &= \frac{g_{2,0} h_0^3}{6} \int d^d x_1 \langle s(x) s(x_1)  \rangle \prod_{i = 1}^3 \int d \tau'_i \langle  s(x_1) s (\tau'_i, \mathbf{0}) \rangle \\
&= \left( \frac{\Gamma \left(\frac{d - 3}{2} \right)}{4 \pi^{\frac{d - 1}{2}}} \right)^3 \frac{g_{2,0} h_0^3}{12 \left(3d - 11 \right)  \left(4 - d\right) \left(\mathbf{x}^2\right)^{(3d-11)/2} } = \frac{g_{2,0} h_0^3}{768 \pi^3 \epsilon |\mathbf{x}|} + \textrm{finite}
\end{split}
\end{equation}
\begin{equation}
\begin{split}
(c) &= - h_0  g_{1,0}^2 \int d^d x_1 d^d x_2 d \tau' \langle s(x) s(x_1)  \rangle \langle s(x_2) s (\tau', \mathbf{0})  \rangle  \langle \bar{\Psi}_i \Psi^i(x_1) \bar{\Psi}_j \Psi^j(x_2) \rangle \\
&= -\frac{   h_0  g_{1,0}^2N_f c_d (d - 2)^2 \Gamma \left(\frac{d}{2} - 1 \right)^4 }{(4 \pi^{d/2})^4} \int  \frac{d^d x_1 d^d x_2 d \tau'}{\left(x_{12} ^2 \right)^{d-1} \left(\left( x - x_1 \right)^2 \right)^{\frac{d}{2} - 1} \left(\mathbf{x}_2^2 + (\tau' - \tau_2)^2 \right)^{\frac{d}{2} - 1} } \\
&= \frac{ h_0  g_{1,0}^2N_f c_d \Gamma \left(\frac{d}{2} - 1 \right)^2 \Gamma \left( d - \frac{7}{2}  \right) }{64 \Gamma \left( d- 2 \right) \pi^{d - \frac{1}{2}} \left(\mathbf{x}^2\right)^{d - \frac{7}{2}}    (4 - d)} = \frac{g_{1,0}^2 h_0 N}{64 \pi^{3} \epsilon |\mathbf{x}|} + \textrm{finite}.
\end{split}
\end{equation}
The bare bulk field may be expressed in terms of the renormalized one as 
\begin{equation}
\langle s_0(x) \rangle = Z_s \langle s(x) \rangle = \left( 1 - \frac{N g_{1,0}^2}{32 \pi^2 \epsilon} + \dots \right) \langle s(x) \rangle ,
\label{Zs-factor}
\end{equation}
where the $Z_s$ factor above is fixed by the renormalization of the bulk 2-point function in the absence of the defect ($Z_s$ determines the anomalous dimension 
of the bulk operator $s(x)$, which is independent of the defect). 
Then, we see that to cancel the divergences and get a finite renormalized one-point function, we need the defect coupling renormalization to be given by
\begin{equation}
\delta h = \frac{g_2 h^3}{192 \pi^2 } + \frac{g_1^2 h N}{32 \pi^2 } + \dots.  
\end{equation}
The beta function of the coupling $h$ may then be computed by requiring that the bare coupling $h_0$ is independent of the scale $M$
\begin{equation}
M \frac{d h_0}{d M} = 0 \implies \frac{\epsilon}{2} \left( h + \frac{\delta h}{\epsilon} \right) + \beta_h \frac{\partial}{\partial h} \left( h + \frac{\delta h}{\epsilon} \right) + \beta_{g_1} \frac{\partial}{\partial g_1} \left( h + \frac{\delta h}{\epsilon} \right) + \beta_{g_2} \frac{\partial}{\partial g_2} \left( h + \frac{\delta h}{\epsilon} \right) = 0.
\end{equation}
This implies 
\begin{equation}
\beta_h = - \frac{\epsilon}{2} h + \frac{g_2 h^3}{96 \pi^2 } + \frac{g_1^2 h N}{32 \pi^2 }.
\end{equation}
So there are two fixed points at $h = 0$ and $h = h_*$ given by 
\begin{equation}
h_*^2 =  \frac{108}{6 -  N + \sqrt{N ^2 + 132 N + 36}}      \xrightarrow{\textrm{Large N}}  \frac{3}{2} + \frac{45}{N}.
\end{equation}
Since the operator $s$ is relevant when $h = 0$, the theory with $h = 0$ is the UV DCFT, while the theory with $h = h_*$ is the IR DCFT. Note that the fixed point value of the defect coupling $h_*$ is of order 1, not a power of $\epsilon$.

We may now calculate the normalized one-point function of $s$. We will normalize it by the two-point function in the usual bulk theory without a defect, which in our conventions is 
\begin{equation}
\langle s (x_1) s(x_2) \rangle = \frac{\mathcal{N}_{s}}{|x_{12}|^{2 \Delta_s}}, \hspace{0.5 cm} \mathcal{N}_{s} = \frac{\Gamma \left(\frac{d}{2} - 1 \right)}{4 \pi^{d/2}} (1 + O(\epsilon)).
\end{equation} 
The normalized one-point function coefficient $a_{s}$ may then be defined by 
\begin{equation} \label{NormOnePointEps}
\langle s (x) \rangle = \frac{\sqrt{\mathcal{N}_{s}} a_{s}  }{|\mathbf{x}|^{\Delta_s}}, \hspace{1cm} a_{s}^2 = \frac{27}{ 6 - N + \sqrt{N ^2 + 132 N + 36}} \xrightarrow{\textrm{Large N}}  \frac{3}{8} + \frac{45}{4N}      .
\end{equation}
To the order we are working at, we can only get the leading $\epsilon^0$ piece for the one-point function.  The order $\epsilon$ piece requires knowing the coupling $h$ to order $\epsilon$, which requires us to go to higher loops in the calculation of beta function.
\subsection{DCFT data}

\subsubsection{Defect operator dimensions: mapping to $H^2 \times S^{d - 2}$ }  \label{SubSec:SpectrumGNY}

We now study the spectrum of defect primaries induced by the bulk scalar and fermions. The operators induced by the bulk fermions are in the fundamental representation of $U(N)$ while the operators induced by the scalar $s$ are $U(N)$ singlets. There is a tower of scalars and fermions on the defect, and it is easiest to see this by performing a Weyl transformation to $H^2 \times S^{d-2}$. Such an approach to obtain DCFT data has been used before in \cite{Kapustin:2005py, Carmi:2018qzm, Herzog:2020lel, Giombi:2020rmc, Giombi:2021uae, Giombi:2021cnr, Cuomo:2021kfm,Sato:2021eqo}.

Recall that the metric of flat space is related to the metric of $H^2 \times S^{d-2}$ by a Weyl transformation
\begin{equation}
ds^2 = \rho^2 \left( \frac{d \rho^2 + d \tau^2}{\rho^2}  + d s^2_{S^{d - 2}}\right) = \rho^2 ds^2_{H^2 \times S^{d - 2}}.
\end{equation}
In this picture, the defect is located at the boundary of $H^2$. The action of the GNY model on this space is 
\begin{equation}
\begin{split}
S &= \int d^d x \sqrt{g} \left( \frac{(\nabla_{\mu} s)^2}{2} + \frac{(d- 2)(d - 4)}{8} s^2 - \left( \bar{\Psi}_i \gamma\cdot \nabla \Psi^i + g_{1,0} s \bar{\Psi}_i \Psi^i\right) + \frac{g_{2,0}}{24} s^4   \right)  \\
& + h_0 \int d \tau s (\tau, \mathbf{0})  .
\end{split}
\end{equation}
We then perform a KK reduction on $S^{d-2}$ to obtain a theory on $H^2$. To do that, we split the gamma matrices into blocks as  \cite{Camporesi:1995fb, Chester:2015wao, Maldacena:2018gjk}
\begin{equation}
\gamma^1 = \sigma^1 \otimes I, \hspace{0.5cm} \gamma^2 = \sigma^2 \otimes I, \hspace{0.5cm} \gamma^i = \sigma^3 \otimes \Gamma^i 
\end{equation} 
where $I$ is the $c_{d - 2}$-dimensional identity, $\Gamma^i$ are gamma matrices in $d - 2$ dimensions and $\sigma^i$ are Pauli matrices. The scalar Laplacian and Dirac operator split as 
\begin{equation}
 (\nabla^2)_{H^2 \times S^{d-2}} = \nabla^2_{H^2} + \nabla^2_{S^{d-2}}, \hspace{1cm}   \left( \slashed{\nabla} \right)_{H^2 \times S^{d-2}} = \slashed{\nabla}_{H^2} \otimes I + \sigma^3 \otimes \slashed{\nabla}_{S^{d-2}} .
\end{equation}
We then expand the scalar and the fermion into the eigenfunctions of the Laplacian and the Dirac operator on $S^{d-2}$
\begin{equation}
\begin{split}
s &= \sum_{l, m}t_{lm} (\rho, \tau) Y_{lm}, \hspace{1cm} \left(\nabla^2 \right)_{S^{d-2}} Y_{lm} = -l (l + d - 3) Y_{lm} \\
\Psi &= \sum_{l, m}\left( \psi^+_{lm} \otimes \chi^+_{lm} + \psi^-_{lm} \otimes \chi^-_{lm} \right), \hspace{1cm} \left( \slashed{\nabla} \right)_{S^{d-2}} \chi^{\pm}_{lm}  = \pm i \left( l + \frac{d}{2} - 1 \right)  \chi^{\pm}_{lm}.
\end{split}
\end{equation}
The sum over $l$ runs from $0$ to infinity, while the sum over $m$ runs over the degeneracy of eigenfunctions of the Laplacian and Dirac operator, respectively. With this reduction, we obtain an effective action on $H^2$. The quadratic terms of this effective action are as follows
\begin{equation}
\begin{split}
S_2 &= \int \frac{d \tau d \rho}{\rho^2} \sum_{l ,m} \bigg[\frac{\nabla_{\mu} t^*_{l,m} \nabla^{\mu} t_{l, m}}{2} + \frac{1}{2} \left(l (l + d - 3) + \frac{(d- 2)(d- 4)}{4} \right) t^*_{l,m} t_{l,m}   \\
& - \sum_{\pm} \left( \bar{\psi}_{lm}^{\pm} \slashed{\nabla}_{H^2} \psi_{lm}^{\pm} \pm i \left( l + \frac{d}{2} - 1 \right) (\bar{\psi}_{lm}^{\pm})^{\dagger} \sigma^3 \psi^{\pm}_{lm}  \right)\bigg].
\end{split}
\end{equation}
We are assuming the following normalizations for the eigenfunctions 
\begin{equation}
\int_{S^2} Y^*_{l,m} Y_{l',m'} = \delta_{ll'} \delta_{mm'}, \hspace{0.5cm} \int_{S^2} {\chi^{\pm}}_{lm} \chi_{l'm'} = \delta_{ll'} \delta_{mm'}.
\end{equation}
In the action above, we have fermions with chiral mass but we can perform a rotation $\psi \rightarrow e^{i \alpha \sigma^3} \psi$. This leaves the kinetic term invariant for all $\alpha$, while for $\alpha = \pi/4$, the mass term transforms as $- i \bar{\psi} \sigma^3 \psi \rightarrow \bar{\psi} \psi$. So we have a tower of scalars and fermions on $H^2$, which give rise to a tower of scalar and fermion operators on the defect. The dimensions of these defect operators are given by the usual mass dimension relations familiar in the AdS/CFT context: 
\begin{equation} \label{DefectSpectrumFree}
\hat{\Delta}_l^s = \frac{d}{2} - 1 + l, \hspace{1cm} \hat{\Delta}_l^f = \frac{d-1}{2} + l  
\end{equation}
for scalars and fermions respectively. 

We may also calculate the corrections to the dimensions of these defect operators. For that purpose, we will use the bulk equations of motion as we now explain (for previous works on using bulk equations of motion to obtain CFT data, see \cite{Rychkov:2015naa, Giombi:2019enr, Giombi:2020xah, Giombi:2021cnr, Bissi:2022bgu}). 

Recall that the two-point function of the bulk scalar with the transverse spin $l$ primary on the defect takes the form \cite{Billo:2016cpy} 
\begin{equation}\label{BulkDefectTwoPoint}
\langle s(x) \hat{s}_l (\tau', \mathbf{w}) \rangle = \frac{(\mathbf{x} \cdot \mathbf{w})^l}{|\mathbf{x}|^{\Delta_s - \hat{\Delta}^s_l + l} (\mathbf{x}^2 + (\tau - \tau')^2)^{\hat{\Delta}^s_l}}
\end{equation}
where $\mathbf{w}$ is a null auxiliary vector living in the embedding space.  We normalize the defect operator such that the coefficient above is one. Acting on the above equation with the bulk Laplacian gives 
\begin{equation}\label{EquationOfMotLHS}
\frac{\nabla^2 \langle s(x) \hat{s}_l (\tau', \mathbf{w}) \rangle}{\langle s(x) \hat{s}_l (\tau', \mathbf{w}) \rangle} = \left[\frac{2 \hat{\Delta}^s_l \left( 2 \Delta_s - d + 2 \right)}{\mathbf{x}^2 + (\tau - \tau')^2} - \frac{ \left(\Delta_s - \hat{\Delta}^s_l + l \right) \left(d - 3 + l - \Delta_s + \hat{\Delta}^s_l \right) }{\mathbf{x}^2} \right]. 
\end{equation}
We then use the fact that the bulk scalar satisfies an equation of motion, which implies the following equation for this two-point function, to leading order in $\epsilon$
\begin{equation}
\nabla^2  \langle s(x) \hat{s}_l (\tau', \mathbf{w}) \rangle = \frac{g_2}{2} \langle s^2 (x) \rangle \langle s(x) \hat{s}_l (\tau', \mathbf{w}) \rangle - g_1^2 \int d^d x_1 \langle \bar{\Psi}_i \Psi^i (x) \bar{\Psi}_j \Psi^j (x_1) \rangle \langle s(x_1) \hat{s}_l (\tau', \mathbf{w}) \rangle.
\end{equation}
Since there are explicit factors of the coupling constant on the R.H.S., we may plug in the correlators in the free theory on R.H.S. and we will get the L.H.S. correct to order $\epsilon$. This tells us that the anomalous dimension of the defect operators must satisfy 
\begin{equation}
\begin{split}
&\frac{(\mathbf{x} \cdot \mathbf{w})^l}{(\mathbf{x}^2 + (\tau - \tau')^2)^{ 1 + l}} \left[\frac{2\gamma_s (2 + 2 l)}{\mathbf{x}^2 + (\tau - \tau')^2} + \frac{\left(\hat{\gamma}^s_l - \gamma_s \right) (1 + 2 l)}{\mathbf{x}^2}  \right]  \\
&= \frac{g_2 h^2}{32 \pi^2} \frac{(\mathbf{x} \cdot \mathbf{w})^l}{ (\mathbf{x}^2)(\mathbf{x}^2 + (\tau - \tau')^2)^{ 1 + l}} - \frac{g_1^2 N \Gamma \left(\frac{d}{2} \right)^2}{4 \pi^d} \int d^d x_1 \frac{(\mathbf{x} \cdot \mathbf{w})^l}{\left((x - x_1)^2 \right)^{d-1} (\mathbf{x}_1^2 + (\tau_1 - \tau')^2)^{ \frac{d}{2} - 1 + l}  }  \\
&= \frac{g_2 h^2}{32 \pi^2} \frac{(\mathbf{x} \cdot \mathbf{w})^l}{ (\mathbf{x}^2)(\mathbf{x}^2 + (\tau - \tau')^2)^{ 1 + l}} + \frac{g_1^2 N (1 + l)}{8 \pi^2} \frac{(\mathbf{x} \cdot \mathbf{w})^l}{ (\mathbf{x}^2 + (\tau - \tau')^2)^{ 2 + l}}
\end{split}
\end{equation} 
where we used that $\mathbf{w}^2 = 0$. This reproduces the correct anomalous dimension of the bulk scalar $\gamma_s = Ng_1^2/(32\pi^2)$ (see \cite{Fei:2016sgs}, and also the $Z_s$ factor in (\ref{Zs-factor})), and tells us that the defect operator of spin $l$ has dimensions 
\begin{equation} \label{DefectDimScEps}
\hat{\gamma}^s_l = \gamma_s + \frac{9 \epsilon}{(N + 6) (1 + 2 l)} \implies \hat{\Delta}^s_l = \Delta_s + l + \frac{9 \epsilon}{(N + 6) (1 + 2 l)} = 1 + l + \frac{6 (1 - l)}{(N + 6) (1 + 2 l)} \epsilon. 
\end{equation}
where we used the known results for the fixed point value of the bulk coupling constants \cite{Fei:2016sgs}. In the limit $l \rightarrow \infty$, the anomalous dimension of the defect primaries is the same as the anomalous dimension of bulk operator, which is consistent with the results of \cite{Lemos:2017vnx}. The $l = 1$ operator should be identified with the displacement and should have protected dimension equal to $2$ to all orders in perturbation theory. This follows from the equation that defines the displacement 
\begin{equation}
\partial_{\mu} T^{\mu i}(x) = D^i (\tau) \delta^{d-1} (\mathbf{x})
\end{equation} 
where $i$ here represents transverse directions. 

This technique of using bulk equations of motion to obtain the anomalous dimensions of defect primaries can also be used in the $O(N)$ model with a localized magnetic field line defect.  We perform this calculation in Appendix \ref{App:EOM} and show its consistency with the computation in \cite{Cuomo:2021kfm} of the scaling dimensions of $l=0$ and $l=1$ defect primaries.

For the leading $l = 0$ operator (let us denote it by $\hat{s}$), there is another way to calculate its scaling dimensions by adapting to the case of a line defect the standard argument that relates the scaling dimension of the perturbing operator to the derivative of the beta function (see for instance \cite{Gubser:2008yx}). To obtain such a relation, note that away from the fixed point, there is a defect ``stress-tensor" $T_D$ localized on the defect \cite{Cuomo:2021rkm}, which in this case is proportional to the beta function of the coupling $h$
\begin{equation}
T_D (\tau) = \beta_h \hat{s} (\tau).
\end{equation}
This defect stress-tensor must have fixed dimension equal to one, hence when we differentiate the above equation with respect to the renormalization scale $M$, we get 
\begin{equation}
\Delta_{\hat{s}} = 1 + \frac{\partial \beta_h}{\partial h}.
\end{equation} 
In the UV DCFT, i.e. $h = 0$, the defect is trivial, so  $\Delta_{\hat{s}}$ is the same as the dimension $\Delta_s$ of the bulk operator $s$ (note that this also implies that in the beta function, the term linear in $h$ must be equal to $\Delta (s) - 1$, which is true). In the IR DCFT, we have 
\begin{equation} \label{LowScEps}
\Delta_{\hat{s}} = 1 + \frac{\partial \beta_h}{\partial h} \bigg|_{h = h_*} = 1 + \frac{6 \epsilon}{N + 6}.
\end{equation}
Note that this agrees with the result (\ref{DefectDimScEps}) obtained above from the equation of motion method, for $l=0$. One can see that the operator $\hat{s}$ is irrelevant in the IR DCFT. Since there is no other candidate for a relevant operator on the defect, this implies that the IR DCFT is stable. 

Next, we make a comment on the anomalous dimension of the defect fermions. We could proceed as we did above for the scalar. However, this involves working out the tensor structures involved in bulk-defect two point function of a fermion. We leave this for a future work, and here just point out how this may be done by the usual Feynman diagram approach. We need to compute the two-point function of the fermion localized on the defect. To leading order in perturbation theory, we get
\begin{equation}
\begin{split}
\langle \bar{\Psi}_i (\tau_1) \Psi^j (\tau_2) \rangle &= \delta_i^j \left[ \frac{\Gamma \left(\frac{d}{2} \right) \gamma^0 (\tau_{12})}{2 \pi^{\frac{d}{2}} |\tau_{12}|^d } - \frac{h g_1 \Gamma \left(\frac{d}{2} \right)^3 }{8 (d-2) \pi^{\frac{ 3 d}{2}} } \int d^d x d \tau \frac{\gamma\cdot (x - \tau_1) \gamma\cdot (x - \tau_2)}{|x - \tau_1|^d |x - \tau_2|^d |x - \tau|^{d - 2}}  \right]
\\
&= \delta_i^j \left[ \frac{\Gamma \left(\frac{d}{2} \right) \gamma^0 (\tau_{12})}{2 \pi^{\frac{d}{2}} |\tau_{12}|^d } - \frac{h g_1 \Gamma \left(\frac{d}{2} \right)^2 \Gamma \left(\frac{d -  3}{2} \right) }{16 \pi^{\frac{ 3 d - 1}{2}} } \int d^d x \frac{\gamma\cdot (x - \tau_1) \gamma\cdot (x - \tau_2)}{|x - \tau_1|^d |x - \tau_2|^d \left(\mathbf{x}^2 \right)^{\frac{d - 3}{2}}}  \right] \\
&= \delta_i^j \left[ \frac{\Gamma \left(\frac{d}{2} \right) \gamma^0 (\tau_{12})}{2 \pi^{\frac{d}{2}} |\tau_{12}|^d } - \frac{h g_1 (4 - d) \Gamma \left(d - \frac{5}{2} \right) \Gamma \left(\frac{5 - d}{2} \right) }{2^{7 - d} \pi^{d - 1} (d-3)  \Gamma \left(3 - \frac{d}{2} \right) (\tau_{12}^2)^{d - \frac{5}{2}}}  \right]
\end{split}
\end{equation}
To go from the second line to the third line, we first did the integral over $x^0$ by introducing a Feynman parameter, then performed the integral over $\mathbf{x}$ by introducing another Feynman parameter, and finally performed the integral over the two Feynman parameters. Note that to obtain the correct spinor structure, we need to project this two-point function onto the defect fermion representation. But it is already clear that the order $g_1$ term actually vanishes in $d = 4$ and does not contribute to the anomalous dimension of defect fermions. So the anomalous dimension must start at order $g_1^2$. The fact that this diagram should not contribute to anomalous dimensions is also clear from a symmetry argument: there is a parity symmetry in the original theory which is broken by the defect. It may be restored if we demand that $h \rightarrow - h $ under the symmetry. So the anomalous dimensions must be even functions of $h$. 

\subsubsection{g-function}
In this subsection, we calculate the defect $g$-function defined as the normalized partition function in the presence of a circular defect of radius $R$
\begin{equation}
\log g = \log \left( Z^{\textrm{bulk $+$ defect}} / Z^{\textrm{bulk}} \right).
\end{equation}
It was shown in \cite{Cuomo:2021rkm} that the defect entropy defined by 
\begin{equation}
s = \left(1 - R \frac{\partial}{\partial R} \right) \log g
\end{equation}
decreases monotonically along the defect RG flows. Using the fact the $\log g$ satisfies the Callan-Symanzik equation 
\begin{equation}
 \left( R \frac{\partial}{\partial R} + \beta_h \frac{\partial}{\partial h} \right) \log g = 0
\end{equation}
we see that at the fixed point, $\log g$ and $s$ must be equal. 

For the trivial defect, $\log g= 0$ and since $h$ generates an RG flow, we know that there must be a non-trivial IR DCFT with $\log g < 0$. In what follows, we will calculate this quantity at the IR fixed point. Let's place the defect on a unit circle around the origin parametrized by $\tau$, $x^{\mu}(\tau) = \left( \cos \tau, \sin \tau, 0, \dots \right)$. There  are three diagrams that we need, to calculate this to leading order in $\epsilon$
\begin{equation}
\log g =\underset{\hbox{$(a)$}}{\vcenter{\hbox{\includegraphics[scale=0.85]{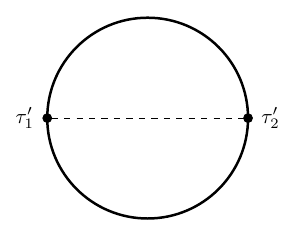}}} }\ \   + \ \  \underset{\hbox{$(b)$}}{\vcenter{\hbox{\includegraphics[scale=0.85]{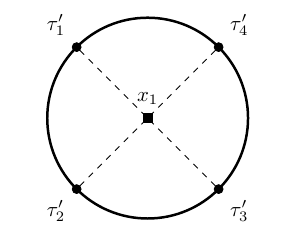}}}} \ \   + \ \  \underset{\hbox{$(c)$}}{\vcenter{\hbox{\includegraphics[scale=0.85]{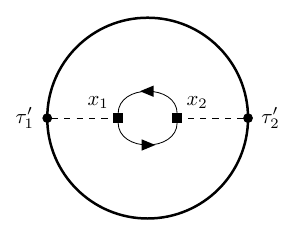}}}}. 
\end{equation}
The diagrams $(a)$ and $(b)$ were already calculated in \cite{Cuomo:2021kfm}. In $d = 4 - \epsilon$, their results are
\begin{equation} 
\begin{split}
(a) &= \frac{h_0^2}{2} \int d \tau_1' d \tau_2' \langle s (x(\tau_1')) s (x(\tau_2'))  \rangle =  - \frac{\epsilon}{8} h_0^2 + O(\epsilon^2), \\
(b) &= \frac{h_0^4 g_{2,0}}{24} \int d^d x_1 \prod_{i = 1}^4 \left( \int d \tau_i'  \langle s (x(\tau_i')) s (x_1)  \rangle \right) =  \frac{h_0^4 g_{2,0}}{384 \pi^2 } + O(\epsilon)  .
\end{split}
\end{equation}
Here we calculate the last one involving a fermion loop 
\begin{equation}
\begin{split}
(c)&= \frac{   h_0^2  g_{1,0}^2 N_f c_d (d - 2)^2 \Gamma \left(\frac{d}{2} - 1 \right)^4 }{2(4 \pi^{d/2})^4} \int \frac{d^d x_1 d^d x_2 d \tau_1' d \tau_2'}{ \left(x_{12} ^2 \right)^{d-1} \left(\left( x(\tau_1') - x_1 \right)^2 \right)^{\frac{d}{2} - 1} \left(\left( x(\tau_2') - x_2 \right)^2 \right)^{\frac{d}{2} - 1} } \\
&= - \frac{   h_0^2  g_{1,0}^2 N  \Gamma \left(\frac{d}{2} - 1 \right)^2 }{128 \pi^d (d - 3) (4 - d) } \int \frac{ d \tau d \tau'}{ \left(\left( x(\tau) - x(\tau') \right)^2 \right)^{d - 3}} \\
&= -\frac{   h_0^2  g_{1,0}^2 N  \Gamma \left(\frac{d}{2} - 1 \right)^2 \Gamma \left(\frac{7 - 2 d}{2} \right) }{2^{2 d - 1} \pi^{d - \frac{3}{2}} (d - 3) \Gamma \left(5 - d\right) }.
\end{split}
\end{equation}
In $d = 4 - \epsilon$, this is 
\begin{equation}
(c) =  \frac{h_0^2  g_{1,0}^2 N }{64 \pi^2} + O(\epsilon).
\end{equation}
Combining all three terms and plugging in the bare couplings in terms of renormalized ones, we get 
\begin{equation}
\log g = - \frac{\epsilon}{8} h^2 + \frac{h^4 g_{2}}{768 \pi^2 } + \frac{h^2  g_{1}^2 N }{128 \pi^2} + O(g_2^2, g_1^4).
\end{equation}
It may be checked that its derivative with respect to $h$ is equal to $\beta_h/2$. The fact that $\partial \log g/ \partial h \propto \beta_h$ is expected on general grounds (see for instance \cite{Beccaria:2017rbe, Cuomo:2021kfm} for related discussions). 

At the fixed point, we finally find
\begin{equation}
\log g \bigg|_{h = h_*} = - \frac{81 \epsilon}{2 (N + 6) \left( 6 -  N + \sqrt{ N ^2 + 132 N + 36} \right) }.
\end{equation}
This is negative, consistent with the $g$-theorem.

\section{Line defect at large $N$} \label{Sec:LargeN}
In this section, we study the line defect in the large $N$ description given by the action in \eqref{ActionLargeN}. As is well known, the $1/N$ perturbation theory in this model may be developed by using the usual propagator for $\Psi$ and the following resummed propagator for $\sigma$ (see for instance \cite{Goykhman:2020tsk})
\begin{equation}
\langle\sigma_0 (x) \sigma_0(0) \rangle = \frac{\mathcal{N}_{\sigma} }{|x|^{2 + \delta}}, \hspace{0.5cm} \mathcal{N}_{\sigma} = - \frac{2^d \sin \left( \frac{\pi d}{2} \right) \Gamma \left(\frac{d - 1}{2} \right)}{\pi^{\frac{3}{2}} \Gamma \left(\frac{d}{2} - 1 \right)}
\end{equation}
where we introduced a regulator $\delta$ which we will set to zero at the end. This makes the dimension of the bare field $1 + \delta/2$ at large $N$, so the bare coupling $h_0$ also acquires a dimension $-\delta/2$. We also need to introduce a coupling $\tilde{g}$ in the bulk with dimension $- \delta/2$
\begin{equation}
S = \int d^d x \left( \bar{\Psi}_i \gamma \cdot \partial \Psi^i + \frac{\tilde{g}_0}{\sqrt{N}} \sigma_0 \bar{\Psi}_i \Psi^i   \right) + h_0 \int d \tau \sigma_0(\tau, \mathbf{0}).
\end{equation} 
Note that this coupling $\tilde{g}$ is really just a rescaling of the $\sigma$ field and in particular, the $\beta$-function of $g$ must vanish at $\delta = 0$
\begin{equation}
\tilde{g}_0 = M^{- \delta/2 }\tilde{g}, \hspace{1cm} \beta_{\tilde{g}} = \frac{\delta}{2} \tilde{g}.
\end{equation} 
We will set the renormalized $\tilde{g} = 1$ at the end. As before, we express the defect bare coupling in terms of the renormalized coupling as 
\begin{equation}
h_0 = M^{-\delta/2}\left(h+\frac{\delta_1 h}{\delta} + \dots\right).
\end{equation}
Following the same procedure as in the GNY case, we calculate the one-point function of $\sigma$ and hence calculate the counterterm $\delta_1h$. The one-point function gets contributions from the following diagrams 

\begin{equation}
\langle \sigma_0(x) \rangle =   \underset{\hbox{$(a)$}}{\vcenter{\hbox{\includegraphics[scale=0.8]{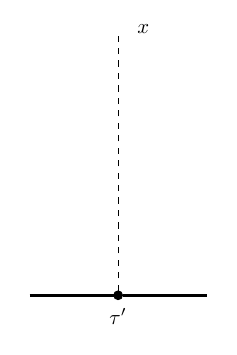}}} }\ \   + \ \  \underset{\hbox{$(b)$}}{\vcenter{\hbox{\includegraphics[scale=0.8]{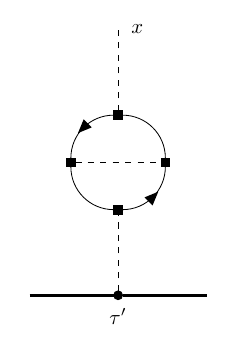}}}} \ \ + \ \  \underset{\hbox{$(b')$}}{\vcenter{\hbox{\includegraphics[scale=0.8]{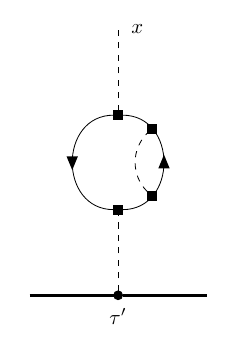}}}}  \ \   + \ \  \underset{\hbox{$(c)$}}{\vcenter{\hbox{\includegraphics[scale=0.8]{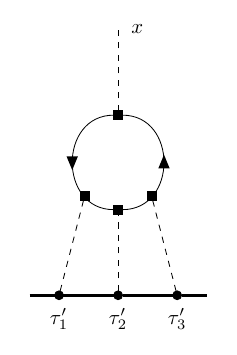}}}}.
\end{equation} 
The first diagram is just the tree level piece 
\begin{equation}
(a) =  -h_0 \int d \tau' \langle\sigma_0 (x) \sigma_0(\tau', \mathbf{0}) \rangle =  - \frac{h_0 \mathcal{N}_{\sigma} \pi }{|\mathbf{x}|}.  
\end{equation}
For the other three loop-diagrams, we will only extract the $1/\delta$ poles. The divergent piece of the combination $(b)+(b')$ has the structure 
\begin{equation}
(b) + (b') = \frac{h_0 \tilde{g}^4 b }{N |\mathbf{x}| \delta}.
\end{equation} 
Notice that this diagram is similar to the one used to calculate the $O(1/N)$ contribution to the $\sigma$ anomalous dimension in the usual theory without defect, except that one of its points is integrated over a line. So we should be able to relate the coefficient $b$ to the bulk anomalous dimension of $\sigma$. To do that, note that if we were to calculate the anomalous dimension of the field $\sigma$ by calculating the divergent piece of the two-point function, we would instead get
\begin{equation}
Z_{\sigma}^2 \langle \sigma(x) \sigma(0) \rangle = \frac{\mathcal{N}_{\sigma} }{|x|^{2 }} - \frac{ \tilde{g}^4 b }{N \pi |x|^{2} \delta}.
\end{equation}
Thus $Z_{\sigma} = 1 -  \frac{\tilde{g}^4 b}{2 N \pi \mathcal{N}_{\sigma} \delta} $ and $\gamma_{\sigma} = \beta_{\tilde{g}} (\partial Z_{\sigma} / \partial \tilde{g} )  $. So we can write $b$ in terms of the anomalous dimension of $\sigma$ 
\begin{equation} \label{LargeNb}
b = - N \pi \mathcal{N}_{\sigma}  \gamma_{\sigma}, \hspace{0.5cm} \gamma_{\sigma} =  \frac{2^{d + 1} (d-1) \sin \left( \frac{\pi d}{2} \right) \Gamma \left(\frac{d - 1}{2} \right)}{N d (d - 2)\pi^{\frac{3}{2}} \Gamma \left(\frac{d}{2} - 1 \right)}
\end{equation}
where we used $\gamma_{\sigma}$ from \cite{Fei:2016sgs}.

Finally, the last diagram involves the following integral 
\begin{equation}  \label{IntegralLargeN}
\begin{split}
(c) &= - \frac{ \tilde{g}_0^4 h_0^3 \mathcal{N}_{\sigma}^4 M^{-4\delta} \Gamma \left(\frac{d}{2} \right)^4 }{16 N \pi^{2 d}} \int \left( \prod_{i = 1}^3 d^d x_i d \tau'_i \frac{1}{|x_i - x (\tau'_i)|^{2 + \delta}} \right) \int d^d x_4 \frac{1}{|x - x_4|^{2 + \delta}} \\
& \times \frac{(x_{14} \cdot x_{1 2}) (x_{23} \cdot x_{34}) - (x_{14} \cdot x_{ 23}) (x_{12} \cdot x_{34}) + (x_{14} \cdot x_{34}) (x_{23} \cdot x_{12}) }{|x_{12}|^d |x_{23 }|^d |x_{3 4}|^d |x_{14 }|^d }.
\end{split}
\end{equation} 
This integral looks very hard to calculate, but we know that the result is supposed to take the form 
\begin{equation}
(c) = \frac{ \tilde{g}_0^4 h_0^3 C(\delta) M^{-4\delta}}{|\mathbf{x}|^{1 + 4 \delta}} \approx \frac{h_0^3 \tilde{g}_0^4 }{|\mathbf{x}|} \left( \frac{c}{ N \delta} + \textrm{finite terms as $\delta \rightarrow 0$}  \right).
\end{equation}
We are only interested in the constant $c$ for the purposes of calculating the $\beta$-function. We calculate this constant in Appendix \ref{App:LargeNInt} and here we just report the result
\begin{equation} \label{LargeNc}
c =  -\frac{\mathcal{N}_{\sigma}^4 \pi^{4}   \Gamma \left(\frac{d - 2 }{2} \right)^2 \left[ (d - 3) \left(\psi \left(\frac{d}{2}-1\right)-\psi (d-3)\right) - 1  \right] }{12  (d - 3) \Gamma(d - 2) \sin \left(\frac{d \pi}{2} \right) } .
\end{equation}
To cancel the $1/\delta$ divergences, the bare coupling should then be 
\begin{equation}
h_0 = M^{-\delta/2}\left(h+\frac{h \tilde{g}^4 b}{2 N \delta \pi \mathcal{N}_{\sigma}} + \frac{h^3 \tilde{g}^4 c}{ N \delta \pi \mathcal{N}_{\sigma}}  + \dots\right).
\end{equation}
Setting $\tilde{g} = 1$, this gives the $\beta$-function for $h$ and the fixed point at leading order in $N$
\begin{equation}
\begin{split}
\beta_{h} &= -\frac{h b}{ N  \pi \mathcal{N}_{\sigma}} - \frac{3h^3 c}{ N  \pi \mathcal{N}_{\sigma}} \\
h_*^2 &=  \frac{2^{3-2 d} (d-3) (d-1) \csc \left(\frac{\pi  d}{2}\right) \Gamma (d-2)}{(d-2) d \left((d-3) H_{\frac{d}{2}-2}+(3-d) H_{d-4}-1\right) \Gamma \left(\frac{d-1}{2}\right)^2}
\end{split}
\end{equation}
where $H_{a}$ is the harmonic number and $b$ and $c$ are given by Eqs. \eqref{LargeNb} and \eqref{LargeNc}, respectively.  This immediately gives us the scaling dimension of the defect operator $\hat{\sigma}$ as 
\begin{equation}
\Delta (\hat{\sigma}) = 1 + \frac{\partial \beta_{h} }{\partial h} \bigg|_{h = h_*} = 1 - \frac{2^{d + 2} (d-1) \sin \left( \frac{\pi d}{2} \right) \Gamma \left(\frac{d - 1}{2} \right)}{N d (d - 2)\pi^{\frac{3}{2}} \Gamma \left(\frac{d}{2} - 1 \right)}.   
\end{equation}
In $d = 4 - \epsilon$, this is 
\begin{equation} \label{LowScN}
\Delta (\hat{\sigma})\bigg|_{d = 4 - \epsilon} = 1 + \frac{6 \epsilon}{N} + O(1/N^2). 
\end{equation}
This is consistent with what we found above using $\epsilon$ expansion techniques for the lowest dimension scalar on the defect, \eqref{LowScEps}. We may also calculate the regularized one-point function of $\sigma$ at leading order in $N$ 
\begin{equation}
\langle \sigma (x) \rangle = \frac{ \sqrt{\mathcal{N}_{\sigma}} a_{\sigma}  }{|\mathbf{x}|^1} = - \frac{h_* \mathcal{N}_{\sigma} \pi }{|\mathbf{x}|} \implies a_{\sigma}^2 = -\frac{(d-3) (d-1)}{(d-2) d \left((d-3) H_{\frac{d}{2}-2}+(3-d) H_{d-4}-1\right)}.
\end{equation}
To compare with GNY results, the $\sigma$ operator in the large $N$ formalism must be identified with $s$ in the GNY formalism, up to normalization. In $d = 4 - \epsilon$, the above result for the normalized one-point function of $\sigma$ is equal to $3/8 + O(\epsilon)$, which is consistent with \eqref{NormOnePointEps}.

Let us now discuss the spectrum of defect operators at large $N$. By the same arguments we gave above in \ref{SubSec:SpectrumGNY}, the spectrum of defect fermions at leading order in large $N$ is still the same as described by \eqref{DefectSpectrumFree}. However, the $\sigma$ operator is no longer a free field even at infinite $N$, so the spectrum of defect scalars differs from \eqref{DefectSpectrumFree}. To obtain this spectrum, note that at leading order in large $N$, the two-point function of $\sigma$ is given by 
\begin{equation}
\langle \sigma (x_1) \sigma(x_2) \rangle = \frac{h_*^2 \mathcal{N}_{\sigma}^2 \pi^2}{|\mathbf{x}_1| |\mathbf{x}_2|}  +  \frac{\mathcal{N}_{\sigma} }{x_{12}^{2}}.
\end{equation}
To obtain the leading large $N$ defect spectrum, we should decompose the above two-point function into defect channel conformal blocks (\cite{Billo:2016cpy, Liendo:2019jpu}). The first term above is the contribution of the identity operator on the defect, while the second term comes from an infinite tower of defect operators with transverse spin $l$ and dimensions $1 + l + 2 m$, as was shown in \cite{Liendo:2019jpu}
\begin{equation}
\frac{1}{x_{12}^{2}} = \frac{1}{|\mathbf{x}_1| |\mathbf{x}_2|} \sum_{m = 0}^{\infty} \sum_{l = 0}^{\infty} b^2_{m,l} \hat{f}_{1 + l + 2 m, l} 
\end{equation}
where $\hat{f}_{1 + l + 2 m, l} $ is the defect channel conformal block for the scalar two-point function. We are suppressing the dependence of the conformal block on the cross-ratios. The coefficient $b^2_{m,l}$ was computed in \cite{Liendo:2019jpu} and is given by 
\begin{equation}
b^2_{m,l} = \frac{\Gamma \left( 2 - \frac{d}{2} + m \right) \Gamma \left( 1 + l +  2m \right)  \Gamma \left( l +  \frac{d-1}{2} \right)  \Gamma \left( \frac{1}{2} + l + m \right) }{ m! \ l! \ \Gamma \left( 2 - \frac{d}{2} \right) \Gamma \left( l +  \frac{d-1}{2} + m \right) \Gamma \left( \frac{1}{2} + l + 2 m \right) }.
\end{equation}
In $d = 4$, the above coefficient is only non-zero for $m = 0$, so only a single operator of dimension $1 + l$ for each transverse spin $l$ survives, which is consistent with what we obtained in \eqref{DefectDimScEps}.

\section{Conclusions} \label{Sec:Conclusions}
In this paper, we defined and studied a line defect in the Gross-Neveu-Yukawa universality class using $\epsilon$-expansion and large $N$ techniques. We provided evidence that there is a non-trivial IR DCFT that this line defect flows to. It should of course be possible to obtain more precise estimates of the DCFT data by going to higher orders in $\epsilon$ expansion in $d = 4 - \epsilon$ dimensions. Another immediate future direction is to develop an $\epsilon$-expansion in $d = 2 + \epsilon$ dimensions using the action in \eqref{ActionGN}. In that case, the line defect must be defined by integrating the fermion bilinear along the line. We make some remarks on this below in Appendix \ref{App:GN}. It would be interesting to develop this expansion for arbitrary $N$ and compare with our large $N$ results. 

Symmetry breaking defects, such as the one we considered in this paper, have recently proved useful in Monte-Carlo simulations near the critical point \cite{Assaad:2013xua, Parisen_Toldin_2017}. Typically, near the phase transition, the value of the order parameter is rather small, and therefore in the usual approach of working with the two-point function of the order parameter, one has to deal with a quantity that is quadratically small. However, in the presence of symmetry breaking defects, the one-point function is also non-zero and hence may be used to infer bulk critical exponents. In \cite{Parisen_Toldin_2017}, this approach was also used to determine scaling dimensions of the defect operators for the pinning field defect in the Ising CFT in $d=3$. It would be very interesting to do such a study for the line defect we studied in the Gross-Neveu CFT and verify our predictions using Monte-Carlo.

\section*{Acknowledgments}
We thank Bendeguz Offertaler for collaboration in the initial stages of this project and for several useful discussions. We also thank Fedor Popov for useful discussions. This research was supported in part by the US NSF under Grant No.~PHY-2209997.

\appendix

\section{Line defect in free fermion theory} \label{App:GN}
In this appendix we discuss the line defect in the free fermion theory defined by
\begin{equation}
S =  \int d^d x\bar{\Psi} \gamma \cdot \partial \Psi  + h_0 \int d \tau \bar{\Psi} \Psi (\tau). 
\end{equation} 
Let's calculate the one-point function of $\langle \bar{\Psi} \Psi \rangle $. Already at this order, we need to sum up infinite number of diagrams. The first few of them look like 
\begin{equation} 
\begin{split}
\langle  \bar{\Psi} \Psi (x) \rangle =&\vcenter{\hbox{\includegraphics[scale=0.88]{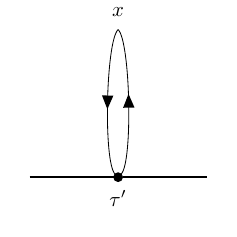}}}    +  \vcenter{\hbox{\includegraphics[scale=0.88]{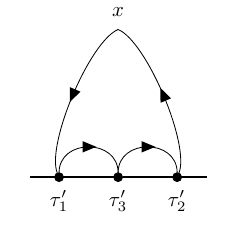}}}     +   \vcenter{\hbox{\includegraphics[scale=0.88]{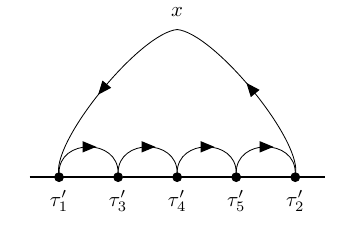}}} + \dots \\
=& -  \frac{ h_0 \Gamma \left(\frac{d}{2} \right)^2 \Gamma \left( d - \frac{3}{2} \right) }{4 \pi^{d - \frac{1}{2}} \Gamma(d - 1) (\mathbf{x}^2)^{\left( d - \frac{3}{2} \right)} } - \frac{ h_0^3  \Gamma \left(\frac{d}{2} \right)^4}{16 \pi^{ 2 d} } \prod_{i = 1}^3 \left( \int d \tau_i' \right) \frac{(x - \tau_1')\cdot (x - \tau_2') \tau_{13}' \tau_{32}'}{(|x - \tau_1'||x - \tau_2'||\tau_{13}'||\tau_{32}'|)^d} \\
& -  \frac{ h_0^5  \Gamma \left(\frac{d}{2} \right)^6}{64 \pi^{ 3 d} } \prod_{i = 1}^5 \left( \int d \tau_i' \right) \frac{(x - \tau_1')\cdot (x - \tau_2') \tau_{13}' \tau_{34}' \tau_{45}' \tau_{52}'}{(|x - \tau_1'||x - \tau_2'||\tau_{13}'||\tau_{34}'| |\tau_{45}'| |\tau_{52}'| )^d} + \dots
\end{split}
\end{equation}
So there are diagrams with increasing powers of $h$ and number of propagators. It is easy to organize them as a geometric progression if we go to momentum space on the line defect (as was done for instance in \cite{Delfino:1994nx})
\begin{equation}
\frac{\Gamma \left(\frac{d}{2} \right) }{2 \pi ^{\frac{d}{2}}} \frac{\gamma^0  \tau_{12}}{|\tau_{12}|^d} = \int \frac{d k }{2 \pi} \frac{i \gamma^0 k \Gamma \left( \frac{3 - d}{2} \right) e^{- i k \tau_{12}} }{2^{d - 1} \pi^{\frac{d - 1}{2} } |k|^{3 - d}}.
\end{equation}
We may then resum the propagator localized on the defect as follows 
\begin{equation}
\begin{split}
& \langle \bar{\Psi} (\tau_1) \Psi (\tau_2) \rangle \\
&= \sum_{n = 0}^{\infty}  (-h_0)^n \left( \frac{\Gamma \left(\frac{d}{2} \right) \gamma^0 }{2 \pi^{\frac{d}{2}}} \right)^{n + 1} \left( \prod_{i = 1}^n d \tau'_{i} \right) \frac{(\tau_2 - \tau_1') (\tau_1' - \tau_2') \dots (\tau_{n - 1}' - \tau_n')(\tau_1 - \tau_n') }{\left(|\tau_2 - \tau_1'| |\tau_1' - \tau_2'| \dots |\tau_{n - 1}' - \tau_n'||\tau_1 - \tau_n'| \right)^d} \\
&= \int \frac{d k }{2 \pi} \frac{ \Gamma \left( \frac{3 - d}{2} \right) e^{- i k \tau_{12}}}{ ( 4 \pi)^{\frac{d - 1}{2}} \left( (k^2)^{2 - d} +  \frac{h_0^2\Gamma \left( \frac{3 - d}{2} \right)^2}{ ( 4 \pi)^{d - 1}} \right) } \left(- \frac{ h_0 \Gamma \left( \frac{3 - d}{2} \right)}{( 4 \pi)^{\frac{d - 1}{2}}} + i |k|^{2 - d} \left( \frac{\gamma^0 k}{|k|}\right) \right).
\end{split}
\end{equation}
The expression for one-point function is then 
\begin{equation}
\begin{split}
\langle  \bar{\Psi}_i \Psi^i (x) \rangle &=  - \frac{ h_0 N \Gamma \left(\frac{d}{2} \right)^2 \Gamma \left( d - \frac{3}{2} \right) }{4 \pi^{d - \frac{1}{2}} \Gamma(d - 1) (\mathbf{x}^2)^{\left( d - \frac{3}{2} \right)} } + \frac{ h_0^3 N \Gamma \left(\frac{d}{2} \right)^2 \Gamma \left( \frac{3 - d}{2} \right)^2 }{2^{2 d} \pi^{2 d - 1} } \int d\tau_1 d \tau_2  \frac{(x - \tau_1)\cdot (x - \tau_2)    }{(|x - \tau_1||x - \tau_2|)^d  } \\
& \times \int \frac{d k }{2 \pi} \frac{e^{- i k \tau_{12}}}{\left( (k^2)^{2 - d} +  \frac{h_0^2\Gamma \left( \frac{3 - d}{2} \right)^2}{( 4 \pi)^{d - 1}} \right)}.
\end{split}
\end{equation}
In $d = 2$, this simplifies to
\begin{equation}
\langle  \bar{\Psi}_i \Psi^i (x) \rangle = - \frac{h_0 N}{4 \pi |\mathbf{x}| \left( 1 + \frac{h_0^2}{4} \right)}.
\end{equation}

In the interacting GN model, the line defect can be described as  
\begin{equation}
S = - \int d^d x \left(  \bar{\Psi}_i \gamma \cdot \partial \Psi^i + \frac{g}{2} \left( \bar{\Psi}_i \Psi^i \right)^2  \right) + h_0 \int d \tau \bar{\Psi}_i \Psi^i (\tau).
\end{equation}
At the fixed point in $d = 2 + \epsilon$, this must flow to the same IR DCFT as the one we studied in the paper. This is because the operator $\bar{\Psi}_i \Psi^i  $ is identified with the operator $\sigma$ in the large $N$ formulation \eqref{ActionLargeN}, and with the operator $s$ in the GNY description \eqref{ActionGNY}. As the free theory calculation discussed above shows, the calculation should involve summing up an infinite set of diagrams with increasing powers of $h$, and hence appears to be non-trivial. We leave the calculation of the properties of this defect in $d = 2 + \epsilon$ dimensions as an interesting future direction. 

\section{Integral for the large $N$ beta function} \label{App:LargeNInt}
In this appendix, we evaluate the integral for diagram $(c)$ relevant for calculating the beta function of the defect coupling at large $N$. The integral in \eqref{IntegralLargeN} is
\begin{equation} 
\begin{split}
(c) &= - \frac{ \tilde{g}_0^4 h_0^3 \mathcal{N}_{\sigma}^4 M^{-4\delta} \Gamma \left(\frac{d}{2} \right)^4 }{16 N \pi^{2 d}} \left( \prod_{i = 1}^3 d^d x_i d \tau'_i \frac{1}{|x_i - x (\tau'_i)|^{2 + \delta}} \right) \int d^d x_4 \frac{1}{|x - x_4|^{2 + \delta}} \\
& \times \frac{(x_{14} \cdot x_{1 2}) (x_{23} \cdot x_{34}) - (x_{14} \cdot x_{ 23}) (x_{12} \cdot x_{34}) + (x_{14} \cdot x_{34}) (x_{23} \cdot x_{12}) }{|x_{12}|^d |x_{23 }|^d |x_{3 4}|^d |x_{14 }|^d }.
\end{split}
\end{equation} 
In momentum space, the integral is 
\begin{equation}
\begin{split}
(c) &= - \frac{ \tilde{g}_0^4 h_0^3 \mathcal{N}_{\sigma}^4 M^{-4\delta} \pi^{2 d} \Gamma \left(\frac{d - 2 - \delta}{2} \right)^4   }{ 2^{-4 d + 8 + 4 \delta} N \Gamma \left(1 + \frac{\delta}{2} \right)^4 } \left( \prod_{i = 1}^3 \frac{d \tau'_i d^d k_i}{(2 \pi)^d} \frac{e^{- i k_i \cdot x(\tau'_i)}}{|k_i|^{d - 2 - \delta}} \right) \frac{e^{i (k_1 + k_2 + k_3) \cdot x}}{|k_1 + k_2 + k_3|^{d - 2 - \delta}} \times  \\
&  \int \frac{d^d k}{(2 \pi)^d} \biggr(  \frac{[k \cdot (k_1 - k) ][(k + k_2)\cdot (k + k_2 + k_3)] - [(k + k_2)\cdot (k_1 - k)][ k \cdot (k + k_2 + k_3) ]}{\left( |k| |k_1 - k| |k + k_2| |k + k_2 + k_3| \right)^2} \\
&+ \frac{[k \cdot (k + k_2)][ (k_1 - k) \cdot (k + k_2 + k_3) ] }{\left( |k| |k_1 - k| |k + k_2| |k + k_2 + k_3| \right)^2} \biggr).
\end{split}
\end{equation}
The integral over $\tau_i'$ may be performed, and it sets the $\tau$ component of the momentum $k_i$ to $0$. As we mentioned in the main text, the result is expected to take the form 
\begin{equation}
(c) = \frac{ \tilde{g}_0^4 h_0^3 C(\delta) M^{-4\delta}}{|\mathbf{x}|^{1 + 4 \delta}} \approx \frac{h_0^3 \tilde{g}_0^4 }{|\mathbf{x}|} \left( \frac{c}{ N \delta} + \textrm{finite terms as $\delta \rightarrow 0$}  \right).
\end{equation}
We are only interested in the constant $c$ for the purposes of calculating the $\beta$ function. To do that, we may perform a Fourier transform to momentum space and then multiply it by a power of momentum and take the limit of $\delta \rightarrow 0$ followed by the limit of zero momentum 
\begin{equation}
\begin{split}
(\tilde{c}) &= |\mathbf{k}|^{d - 2 - \delta} \int d^{d - 1} \mathbf{x}  \frac{\tilde{g}_0^4 h_0^3 C(\delta) M^{-4\delta}}{|\mathbf{x}|^{1 + 4 \delta}} e^{i \mathbf{k} \cdot \mathbf{x}} \bigg|_{\mathbf{k} \rightarrow 0} \\
&\approx  2^{d - 2} \pi^{\frac{d}{2} - 1} \Gamma \left( \frac{d}{2} - 1 \right) \frac{h_0^3 \tilde{g}_0^4 c}{N \delta} + \textrm{finite terms as $\delta \rightarrow 0$}  .
\end{split}
\end{equation}
Since the $1/\delta$ piece is independent of $\mathbf{k}$, when we do these operations to the integral above, we may first take the zero momentum limit, and then take $\delta \rightarrow 0$ at the end. This gives 
\begin{equation}\label{eq:tildec}
\begin{split}
(\tilde{c}) &= - \frac{\tilde{g}_0^4 h_0^3 \mathcal{N}_{\sigma}^4 M^{-4\delta} \pi^{2 d} \Gamma \left(\frac{d - 2 - \delta}{2} \right)^4  }{ 2^{-4 d + 9 + 4 \delta} N \Gamma \left(1 + \frac{\delta}{2} \right)^4 } \int \frac{d^{d-1} \mathbf{k}_1}{(2 \pi)^{d-1}}  \frac{d^{d-1} \mathbf{k}_1}{(2 \pi)^{d-1}} \frac{1}{\left( |\mathbf{k}_1| |\mathbf{k}_2| |\mathbf{k}_1 + \mathbf{k}_2| \right)^{d - 2 - \delta}}   
\\ &\times \int \frac{d^d k}{(2 \pi)^d} \frac { \left( k_2^2 - k^2 - (k + k_2)^2 \right)}{k^2 (k - k_1)^2 (k + k_2)^2} = (\tilde{c})_1 + (\tilde{c})_2 + (\tilde{c})_3 .
\end{split}
\end{equation}
with the understanding that $k_i^0 = 0$. In the second line, we split the three terms in the numerator, and we will deal with them separately. 
The result of the first integral over $k$ can be represented in terms of a Mellin-Barnes contour integral \cite{Davydychev:1995mq}
\begin{equation}
\begin{split}
&(\tilde{c})_1 =  - \frac{\tilde{g}_0^4 h_0^3 \mathcal{N}_{\sigma}^4 M^{-4\delta} \pi^{3d/2} \Gamma \left(\frac{d - 2 - \delta}{2} \right)^4 c_d  }{ 2^{-3 d + 9 + 4 \delta} N \Gamma \left(1 + \frac{\delta}{2} \right)^4 \Gamma(d - 3) } \int \frac{d^{d-1} \mathbf{k}_1}{(2 \pi)^{d-1}}  \frac{d^{d-1} \mathbf{k}_2}{(2 \pi)^{d-1}} \frac{1}{\left( |\mathbf{k}_1| |\mathbf{k}_1 + \mathbf{k}_2| \right)^{d - 2 - \delta} |\mathbf{k}_2|^{d - 4 - \delta} } \times  \\ 
& \int_{- i \infty}^{i \infty} \int_{- i \infty}^{i \infty} \frac{ds dt}{(2 \pi i )^2} \frac{\Gamma(-s) \Gamma(-t)  \Gamma \left(\frac{d }{2} - 2 - s \right) \Gamma \left(\frac{d }{2} - 2 - t \right) \Gamma \left(1 + s +  t \right) \Gamma \left(3 - \frac{d}{2} + s +  t \right) }{|\mathbf{k}_1|^{- 2 s} |\mathbf{k}_2|^{- 2 t} |\mathbf{k}_1 + \mathbf{k}_2|^{2 s + 2 t + 6 - d } }.
\end{split}
\end{equation}
The integral over $\mathbf{k}_1$ and $\mathbf{k}_2$ may then be performed easily and it gives 
\begin{equation}
\begin{split}
&(\tilde{c})_1 = \frac{\tilde{g}_0^4 h_0^3 \mathcal{N}_{\sigma}^4 \pi^{\frac{d }{2} + 1} 2^{d - 6} \Gamma \left(\frac{d - 2}{2} \right)^4  }{3 N \delta \Gamma \left(\frac{d - 1}{2} \right) \Gamma(d - 3) }  \times \\
& \int_{- i \infty}^{i \infty} \int_{- i \infty}^{i \infty} \frac{ds dt}{(2 \pi i )^2 } \frac{\Gamma(-s) \Gamma(-t)  \Gamma \left(s + \frac{1}{2} \right) \Gamma \left(t + \frac{3}{2} \right) \Gamma \left(\frac{d - 5}{2} - s - t \right) \Gamma \left(3 - \frac{d}{2} + s +  t \right) }{\left(\frac{d }{2} - 2 - s \right) (1 + s + t) } \\
&= \frac{\tilde{g}_0^4 h_0^3 \mathcal{N}_{\sigma}^4 \pi^{\frac{d }{2} + 1} 2^{d - 6} \Gamma \left(\frac{d - 2}{2} \right)^4  }{3 N \delta \Gamma \left(\frac{d - 1}{2} \right) \Gamma(d - 3)  }   \int_{- i \infty}^{i \infty} \frac{dt}{2 \pi i }  \frac{ \Gamma(-t)  \Gamma \left(t + \frac{3}{2} \right)}{\left(\frac{d}{2} - 1 + t \right)} \times  \\
& \int_{- i \infty}^{i \infty} \frac{ds}{2 \pi i} \Gamma(-s)  \Gamma (s + \tfrac{1}{2} ) \Gamma (3 - \tfrac{d}{2} + s +  t ) \Gamma (\tfrac{d - 5}{2} - s - t ) \left( \frac{\Gamma \left( 1  + s +  t \right)}{\Gamma \left(2 + s +  t \right)} - \frac{\Gamma \left( 2-  \frac{ d}{2} + s \right)}{\Gamma \left(3 - \frac{d}{2} + s \right)}  \right).
\end{split}
\end{equation}
In the third and fourth line, we write it in a form so that we can directly do the $s$ integral by applying Barnes' second Lemma to both the terms above. But we also need to add the contribution of the pole at $s = d/2 - 2$, which is not included by the result of Barnes' Lemma. Adding all the pieces, this gives 
\begin{equation}
\begin{split}
&(\tilde{c})_1 = \frac{\tilde{g}_0^4 h_0^3 \mathcal{N}_{\sigma}^4 \pi^{\frac{d }{2} + 2} 2^{d - 6} \Gamma \left(\frac{d - 2}{2} \right)^4  }{3 N \delta \Gamma \left(\frac{d - 1}{2} \right) \Gamma(d - 3) (2 \pi i ) }   \int_{- i \infty}^{i \infty} dt \bigg[  \frac{2 \Gamma \left( 2 - \frac{d}{2} \right) \Gamma \left(\frac{d - 3}{2} \right) \Gamma(2 t) \Gamma(1 - 2 t)}{\left(\frac{d}{2} - 1 + t \right)} \\
&+  \frac{ \Gamma \left( 3 - \frac{d}{2} + t \right)  \Gamma \left( \frac{d}{2} - 2 - t \right)}{\left(\frac{d}{2} - 1 + t \right)}  \left( \frac{ \Gamma(-t) \Gamma(1 + t) \Gamma \left(\frac{d - 3}{2} \right) }{\Gamma \left(\frac{d}{2} - 1 \right)} - \frac{ \Gamma \left(t + \frac{3}{2} \right) \Gamma \left(-t - \frac{1}{2} \right) \Gamma \left( 2 - \frac{d}{2} \right) }{\Gamma \left(\frac{5 -d}{2}\right)} \right)  \bigg] \\
&= - \frac{\tilde{g}_0^4 h_0^3 \mathcal{N}_{\sigma}^4 \pi^{\frac{d }{2} + 5} 2^{d - 5} \Gamma \left(\frac{d - 2}{2} \right)^3  }{3 N \delta \Gamma \left(\frac{d - 1}{2} \right) \Gamma(d - 3) \Gamma \left(\frac{5 - d}{2} \right) (2 \pi i ) } \int_{- i \infty}^{i \infty} dt \frac{\csc (\pi  t ) \csc \left( \pi  \left( \frac{d}{2} - t \right) \right) \sec \left(\frac{d \pi}{2} \right) }{\left(\frac{d}{2} - 1 + t \right)}   .
\end{split}
\end{equation}
We can then sum up all the poles at $t =  m$ and at $ d/2 - 2 + m$ to get 
\begin{equation}
\begin{split}
(\tilde{c})_1 =  -\frac{\tilde{g}_0^4 h_0^3 \mathcal{N}_{\sigma}^4 \pi^{\frac{d }{2} + 4} 2^{d - 4} \Gamma \left(\frac{d - 2}{2} \right)^3 \csc (\pi  d) \left(\psi \left(\frac{d}{2}-1\right)-\psi (d-3)\right) }{3 N \delta \Gamma \left(\frac{d - 1}{2} \right) \Gamma \left(\frac{5 - d}{2} \right)  \Gamma(d - 3) } . 
\end{split}
\end{equation}
Next, we look at the second term, which is easier
\begin{equation}
\begin{split}
(\tilde{c})_2 &= \frac{\tilde{g}_0^4 h_0^3 \mathcal{N}_{\sigma}^4 M^{-4\delta} \pi^{2 d} \Gamma \left(\frac{d - 2 - \delta}{2} \right)^4   }{ 2^{-4 d + 9 + 4 \delta} N \Gamma \left(1 + \frac{\delta}{2} \right)^4 } \\
& \times \int \frac{d^{d-1} \mathbf{k}_1}{(2 \pi)^{d-1}}  \frac{d^{d-1} \mathbf{k}_1}{(2 \pi)^{d-1}} \frac{d^d k}{(2 \pi)^d}  \frac{1}{\left( |\mathbf{k}_1| |\mathbf{k}_2| |\mathbf{k}_1 + \mathbf{k}_2| \right)^{d - 2 - \delta} (k - k_1)^2 (k + k_2)^2 } \\
\end{split}
\end{equation}
Upon integrating $k$ and $\mathbf{k}_1$, one gets
\begin{equation}
\begin{split}
(\tilde{c})_2 &=- \frac{\tilde{g}_0^4 h_0^3 \mathcal{N}_{\sigma}^4 M^{-4\delta} \pi^{2 d} \Gamma \left(\frac{d - 2 - \delta}{2} \right)^4   }{ 2^{-4 d + 9 + 4 \delta} N \Gamma \left(1 + \frac{\delta}{2} \right)^4 } \frac{\pi ^{\frac{3}{2}-d} 2^{-3 d-\delta +5} \sin \left(\frac{\pi  \delta }{2}\right) \csc \left(\frac{\pi  d}{2}\right) \Gamma \left(\frac{1}{2}-\delta \right) \Gamma (\delta ) \Gamma\left(\frac{1}{2} (d+\delta -3)\right)}{\Gamma \left(\frac{d-1}{2}\right) \Gamma \left(\frac{1}{2}(d-\delta -2)\right) \Gamma \left(\frac{d}{2}+\delta -1\right)} \\
&\times \int \frac{d^{d-1}\mathbf{k}_2}{(2\pi)^{d-1}} \frac{1}{|\mathbf{k}_2|^{d-1-3\delta}}
\end{split}
\end{equation}
which is both UV and IR divergent. We introduce a regulator $\mu$ to separate the IR divergence, which gives 
\begin{equation}
\begin{split}
\int \frac{d^{d-1}\mathbf{k}_2}{(2\pi)^{d-1}} \frac{1}{(\mathbf{k}_2^2 + \mu)^{d/2-1/2-3\delta/2}} &= \frac{2^{1-d} \pi ^{\frac{d-1}{2}-d+1} \mu ^{3 \delta /2} \Gamma \left(-\frac{3 \delta }{2}\right)}{\Gamma \left(\frac{1}{2} (d-3 \delta -1)\right)} \\
&= -\frac{2^{2-d} \pi ^{\frac{1}{2}-\frac{d}{2}}}{3 \delta  \Gamma \left(\frac{d-1}{2}\right)} -\frac{2^{1-d} \pi ^{\frac{1}{2}-\frac{d}{2}} \left(\psi ^{(0)}\left(\frac{d-1}{2}\right)+\log (\mu )+\gamma \right)}{\Gamma \left(\frac{d-1}{2}\right)} + O(\delta). 
\end{split}
\end{equation}
Because we only need to extract the UV divergent piece for the purpose of computing the beta function, we discard the $O(1)$ piece containing $\log(\mu)$, so the result for $(\tilde{c})_2$ is
\begin{equation}
(\tilde{c})_2 = -\frac{ \tilde{g}_0^4 h_0^3  \mathcal{N}_{\sigma}^4 \pi^{\frac{d}{2} + 2}  2^{d - 5}  \Gamma \left(\frac{d - 2 }{2} \right)^4 \Gamma \left(2 - \frac{d}{2} \right)}{3 (d - 3) \Gamma(d - 2) N \delta}.
\end{equation}
The third piece in \ref{eq:tildec} gives the same result:
\begin{equation}
(\tilde{c})_3 = -\frac{ \tilde{g}_0^4 h_0^3  \mathcal{N}_{\sigma}^4 \pi^{\frac{d}{2} + 2}  2^{d - 5}  \Gamma \left(\frac{d - 2 }{2} \right)^4 \Gamma \left(2 - \frac{d}{2} \right)}{3 (d - 3) \Gamma(d - 2) N \delta}.
\end{equation}
Combining all the pieces, we get
\begin{equation} 
\begin{split}
c &=  -\frac{\mathcal{N}_{\sigma}^4 \pi^{4}   \Gamma \left(\frac{d - 2 }{2} \right)^2 \left[ (d - 3) \left(\psi \left(\frac{d}{2}-1\right)-\psi (d-3)\right) - 1  \right] }{12  (d - 3) \Gamma(d - 2) \sin \left(\frac{d \pi}{2} \right) } .
\end{split}
\end{equation}

\section{Equations of motion method for line defect in $O(N)$ model}  \label{App:EOM}
Similar to what we did for GNY in the main text, we can also use the bulk equations of motion to determine the anomalous dimensions of spinning defect operators in the Wilson-Fisher $O(N)$ model with a localized magnetic field, which was studied in \cite{Cuomo:2021kfm}.  The action for that model is 
\begin{equation}
S = \int d^d x \left( \frac{1}{2} (\partial_\mu \phi^I)^2 + \frac{\lambda}{4!} (\phi^I\phi^I)^2 \right)+ h \int d\tau \phi^1 (\tau, \mathbf{0}).
\end{equation}
where $I = 1, ..., N$. The bulk theory has a fixed point in $d=4-\epsilon$ at $\lambda_* = \frac{3(4\pi)^2 \epsilon}{N+8}$. The two-point function of the bulk scalar with the transverse spin $l$ primary on the defect takes essentially the same form as \eqref{BulkDefectTwoPoint}:
\begin{equation}
\langle \phi^I(x) \hat{\phi}^J_l (\tau', \mathbf{w}) \rangle = \frac{\delta^{IJ} (\mathbf{x} \cdot \mathbf{w})^l}{|\mathbf{x}|^{\Delta^{\phi} - \hat{\Delta}^{\phi}_l + l} (\mathbf{x}^2 + (\tau - \tau')^2)^{\hat{\Delta}^{\phi}_l}}.
\end{equation}
As before, acting on the above equation with the bulk Laplacian gives \eqref{EquationOfMotLHS}
\begin{equation} \label{ONEquationOfMotLHS}
\frac{\nabla^2 \langle \phi^I(x) \hat{\phi}^J_l (\tau', \mathbf{w}) \rangle}{\langle \phi^I(x) \hat{\phi}^J_l (\tau', \mathbf{w}) \rangle} = \left[\frac{2 \hat{\Delta}^\phi_l \left( 2 \Delta^\phi - d + 2 \right)}{\mathbf{x}^2 + (\tau - \tau')^2} - \frac{ \left(\Delta^\phi - \hat{\Delta}^\phi_l + l \right) \left(d - 3 + l - \Delta^\phi + \hat{\Delta}^\phi_l \right) }{\mathbf{x}^2} \right]. 
\end{equation}
On the other hand, the fact that the bulk scalars satisfy an equation of motion implies that, to leading order in $\epsilon$, we have
\begin{equation}
\begin{split} \label{ONEquationOfMotRHS}
\nabla^2  \langle \phi^I(x) \hat{\phi}^J_l (\tau', \mathbf{w}) \rangle &= \frac{\lambda}{6}\langle \phi^I \phi^K \phi^K (x) \hat{\phi}^J_l (\tau', \mathbf{w}) \rangle\\
&=\frac{\lambda}{6}(1+2\delta^J_1)\frac{ A_{\phi^2}}{|\mathbf{x}|^{d-2}} \langle \phi^I(x) \hat{\phi}^J_l (\tau', \mathbf{w}) \rangle, \qquad A_{\phi^2} = \tfrac{\Gamma(d/2)}{(d-2)2\pi^{d/2}}\tfrac{N+8}{4} + O(\epsilon).
\end{split}
\end{equation}
where $A_{\phi^2}$ is the coefficient of the one-point function of $\phi^2$ in the presence of the defect. This quantity was computed in \cite{Cuomo:2021kfm}, as well as the defect coupling at the fixed point $h_*^2 = N+8 + O(\epsilon)$. Noting that $\gamma^\phi = O(\epsilon^2)$, we set \eqref{ONEquationOfMotRHS} equal to \eqref{ONEquationOfMotLHS} and obtain
\begin{equation}
\hat{\gamma}_l^{\phi_J} = \frac{(1+2\delta^J_1)\epsilon}{2(1+2l)} + O(\epsilon^2).
\end{equation}
Thus the scaling dimensions of the defect operators are
\begin{equation}
 \hat{\Delta}_l^{\phi_J } = \biggr\{
 \begin{array}{l r}
 \Delta^\phi + l +\frac{3\epsilon}{2(1+2l)} \underset{d=4-\epsilon}{=} 1+l +\frac{1-l}{1+2l}\epsilon & J = 1 \\
 \Delta^\phi + l +\frac{\epsilon}{2(1+2l)} \underset{d=4-\epsilon}{=} 1+l -\frac{l}{1+2l}\epsilon & \qquad J = 2,\dots, N
 \end{array}
\end{equation}
Note that the scaling dimensions for $\hat{\Delta}_{l=0}^{\phi_1 }$,  $ \hat{\Delta}_{l=1}^{\phi_1 }$,  $ \hat{\Delta}_{l=0}^{\phi_J }$, and  $ \hat{\Delta}_{l=1}^{\phi_J }$ are consistent with the respective scaling dimensions of $\Delta(\hat{\phi}_1)$, $\Delta(\nabla \hat{\phi}_1)$, $\Delta(\hat{\phi}_a)$, and $\Delta(\nabla \hat{\phi}_a)$ obtained in \cite{Cuomo:2021kfm}, as expected. But this method gives us results for all $l$.

\bibliographystyle{ssg}
\bibliography{LineDefectsDraft-bib}

\end{document}